\documentclass[11pt]{article}

\usepackage[final]{acl}

\usepackage{algorithm}
\usepackage{algpseudocode}
\usepackage{times}
\usepackage{latexsym}
\usepackage{amsmath}
\usepackage{amssymb}
\usepackage{mathtools}
\usepackage{amsthm}
\usepackage{booktabs}
\usepackage{multirow}
\usepackage{tikz}
\usepackage[T1]{fontenc}
\theoremstyle{plain}

\theoremstyle{definition}

\theoremstyle{remark}

\usepackage[utf8]{inputenc}
\usepackage[normalem]{ulem}
\usepackage[capitalize,noabbrev]{cleveref}
\usepackage{microtype}
\usepackage{hyperref}
\usepackage{tabularx}
\usepackage{microtype}
\usepackage{subcaption}
\usepackage{booktabs} 
\usepackage{tcolorbox}
\usepackage{inconsolata}

\usepackage{graphicx}

\usepackage[textsize=tiny]{todonotes}
\definecolor{TokenGreen}{RGB}{162, 217, 137}

\title{Chain-of-Thought Compression Should Not Be Blind: V-Skip for Efficient Multimodal Reasoning via Dual-Path Anchoring}

\author{
    \textbf{Dongxu Zhang}\textsuperscript{1,$\ast$} \quad
    \textbf{Yiding Sun}\textsuperscript{1,$\ast$} \quad
    \textbf{Cheng Tan}\textsuperscript{3,$\ast$} \quad
    \textbf{Wenbiao Yan}\textsuperscript{4} \\
    \textbf{Ning Yang}\textsuperscript{2,$\dagger$} \quad
    \textbf{Jihua Zhu}\textsuperscript{1,$\dagger$} \quad
    \textbf{Haijun Zhang}\textsuperscript{5} \\
    \\
    \textsuperscript{1}School of Software Engineering, Xi'an Jiaotong University \\
    \textsuperscript{2}Institute of Automation, Chinese Academy of Sciences \\
    \textsuperscript{3}Shanghai Artificial Intelligence Laboratory \\
    \textsuperscript{4}Harbin Institute of Technology, Shenzhen \\
    \textsuperscript{5}University of Science and Technology Beijing
}

\begin{document}
\maketitle
{
  \renewcommand{\thefootnote}{\fnsymbol{footnote}}
  \footnotetext[1]{Equal contribution.}
  \footnotetext[2]{Corresponding authors.}
}
\begin{abstract}

While Chain-of-Thought (CoT) reasoning significantly enhances the performance of Multimodal Large Language Models (MLLMs), its autoregressive nature incurs prohibitive latency constraints. Current efforts to mitigate this via token compression often fail by blindly applying text-centric metrics to multimodal contexts. We identify a critical failure mode termed Visual Amnesia, where linguistically redundant tokens are erroneously pruned, leading to hallucinations. To address this, we introduce V-Skip that reformulates token pruning as a Visual-Anchored Information Bottleneck (VA-IB) optimization problem. V-Skip employs a dual-path gating mechanism that weighs token importance through both linguistic surprisal and cross-modal attention flow, effectively rescuing visually salient anchors. Extensive experiments on Qwen2-VL and Llama-3.2 families demonstrate that V-Skip achieves a $2.9\times$ speedup with negligible accuracy loss. Specifically, it preserves fine-grained visual details, outperforming other baselines over 30\% on the DocVQA.

\end{abstract}

\section{Introduction}

The integration of vision and language has catalyzed a paradigm shift in Artificial Intelligence, culminating in the emergence of Multimodal Large Language Models (MLLMs)~\cite{wang2025multimodal}. By extending the Chain-of-Thought (CoT) reasoning capability from text to vision, models such as LLaVA~\cite{li2024llava} and Qwen-VL~\cite{wang2024qwen2} have demonstrated remarkable proficiency in decomposing complex visual queries into step-by-step rationales. This reasoning process significantly enhances performance and interpretability.

\begin{figure*}[t]
  \centering
  \includegraphics[width=0.75\linewidth]{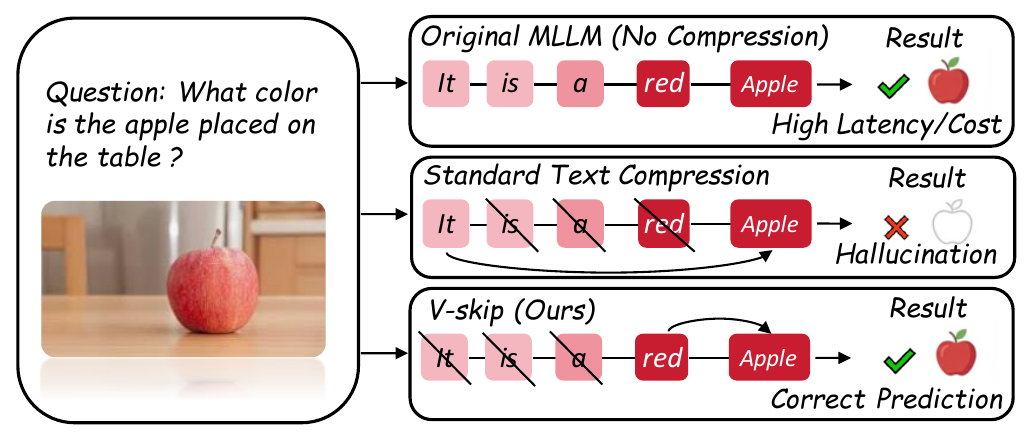}
\caption{Comparison of token compression paradigms.
    \textbf{(Top)} The original MLLMs retains all tokens, ensuring accuracy but incurring high computational costs. 
    \textbf{(Middle)} Standard text-centric compression aggressively prunes the adjective red due to its high linguistic probability given apple, detaching the reasoning from the visual ground truth and leading to hallucination. 
    \textbf{(Bottom)} V-Skip utilizes a dual-path scoring mechanism to identify and rescue visually salient tokens while compressing redundant linguistic fillers.}
  \label{figure1}
\end{figure*}

However, because of the autoregressive nature of Transformer architectures, generating extensive reasoning chains introduces severe latency and computational overhead~\cite{shazeer2019fast,dao2022flashattention}. The linear growth of the Key-Value (KV) cache restricts batch sizes and limits throughput, rendering long chain visual reasoning impractical~\cite{kwon2023efficient}. Empirically, these generated sequences often contain linguistic fillers~\cite{jiang2023llmlingua} that support syntactic fluency yet offer minimal information gain for the final prediction.

To mitigate this inefficiency, token compression techniques such as TokenSkip~\cite{xia2025tokenskip} or LLMLingua-2~\cite{pan2024llmlingua} have been proposed. While effective in text contexts, we argue that directly applying these methods to multimodal tasks precipitates a critical failure mode we term Visual Amnesia. This phenomenon severs the connection to the input image and induces object hallucinations~\cite{li2023evaluating}. 
Text-centric pruning relies on linguistic surprisal. If a token is statistically predictable from the preceding text, it is deemed redundant. However, in multimodal reasoning, token importance is bimodal. For instance, a token like red may be linguistically predictable given apple, yet it serves as an essential visual anchor. As illustrated in Figure~\ref{figure1}, standard algorithms rely solely on linguistic priors and discard such anchors without considering visual context, severing the connection to the input image and inducing object hallucinations~\cite{rohrbach2018object}.

In this paper, we introduce V-Skip, a novel method designed to compress multimodal CoT sequences while strictly preserving visual grounding. We reformulate the compression task as a Visual-Anchored Information Bottleneck (VA-IB) optimization problem. Unlike existing methods that rely solely on language probability~\cite{zhang2025ascot}, V-Skip employs a dual-path scoring mechanism that interrogates both linguistic surprisal and cross-modal attention flow. This allows the model to identify and rescue tokens that are linguistically redundant but visually salient. Furthermore, to avoid the latency overhead of computing attention maps, we treat compression not as an online search but as an offline policy learning problem. We distill the V-Skip pruning policy into the base model via Low-Rank Adaptation (LoRA)~\cite{hu2022lora}, resulting in an efficient reasoner that generates concise, visually grounded rationales without requiring explicit scoring or online filtering during inference.

Our contributions are summarized as follows:

\begin{itemize} 
\item We identify Visual Amnesia as a fundamental failure mode in multimodal CoT compression, where text-centric pruning inadvertently discards tokens essential for visual grounding. 
\item The proposed VA-IB framework reformulates multimodal compression as an information theoretic objective to balance linguistic efficiency with cross-modal grounding.
\item We introduce V-Skip, a novel framework utilizing a dual-path scoring mechanism to selectively preserve visual anchors, distilled into a lightweight adapter for efficient inference.
\item Extensive experiments on Qwen2-VL demonstrate that V-Skip achieves a $2.9\times$ speedup with minimal accuracy trade-offs. It outperforms other baselines over 30\% on DocVQA.
\end{itemize}

\section{Related Work}

\begin{figure*}[t]
  \centering
  \includegraphics[width=1.0\linewidth]{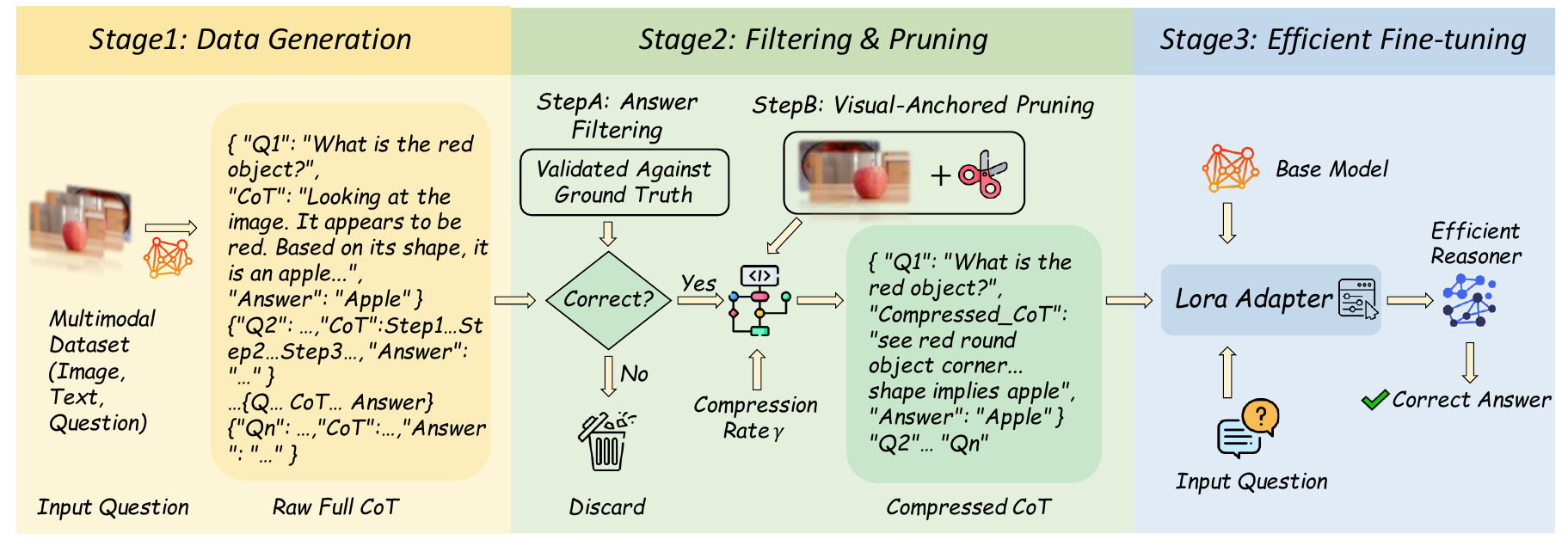}
  \caption{The V-Skip Training Pipeline. Our method automates the construction of efficient multimodal reasoners through three distinct stages. (1) Data Generation, a VLM generates comprehensive CoT rationales for input image-question pairs. (2) Filtering \& Pruning, Step A filters out rationales that lead to incorrect answers compared to the ground truth. Step B applies our Visual-Anchored Pruning mechanism (detailed in Figure~\ref{figure3}), which selectively removes redundant tokens based on a target compression rate $\gamma$, producing concise Compressed\_CoT sequences. (3) Efficient Fine-tuning, the resulting high-quality compressed dataset is used to supervise a Base Model via LoRA, yielding an Efficient Reasoner capable of generating succinct and visually grounded reasoning paths directly.}
  \label{figure2}
\end{figure*}

\subsection{Multimodal Large Language Models}

Recent advancements in MLLMs have bridged the gap between visual perception and linguistic intelligence. Foundational models such as Llama~\cite{dubey2024llama}, LLaVA~\cite{li2024llava}, and Qwen-VL~\cite{wang2024qwen2} project visual features into the embedding space of LLMs, enabling them to process interleaved image-text inputs.

On one hand, text-centric methods like Tree-of-Thoughts~\cite{yao2023tree} and Graph-of-Thoughts~\cite{besta2024graph} explore non-linear reasoning topologies. Parallel to these explorations, multimodal strategies focus heavily on cross-modal alignment and rationale quality. For instance, PICa~\cite{yang2022empirical} converts visual information into captions to prompt GPT-3 for reasoning. Multimodal-CoT~\cite{zhang2023multimodal} adopts a two-stage method that fine tunes models using human annotated rationales. Other works like PromptCoT~\cite{yao2024promptcot} and T-SciQ~\cite{wang2024t} leverage prompt engineering and teacher signals to guide the generation of reasoning paths.

Despite these improvements in accuracy, the resulting extensive reasoning chains often introduce significant computational overhead~\cite{zhang2023multimodal}. In contrast to these generation centric methods, our work focuses on the efficient utilization of reasoning chains, pruning redundant computations while preserving the accurate prediction.

\subsection{Efficient LLM Reasoning \& Compression}
Efforts to mitigate the computational demands of LLMs have focused on system level optimizations. Techniques such as quantization~\cite{dettmers2022gpt3}, structured compression~\cite{zhu2025breaking}, and KV cache optimization strategies like PagedAttention~\cite{kwon2023efficient} effectively reduce memory footprints. However, these methods do not address the fundamental bottleneck of sequence length in autoregressive generation.

To directly tackle the computational complexity, token pruning and compression have emerged as effective strategies~\cite{chuang2024learning,yang2025token}. Methods like TokenSkip~\cite{xia2025tokenskip} accelerate inference by structurally bypassing layers for redundant tokens. Methods like ASCoT~\cite{zhang2025ascot} and LLMLingua-2~\cite{pan2024llmlingua} leverage perplexity and entropy distributions to identify and prune tokens with low linguistic information gain, reducing the generation of unnecessary intermediate steps.

Despite their efficacy in NLP, compression metrics based on linguistic probability often discard critical visual anchors. V-Skip distinguishes itself as the first framework to explicitly incorporate Visual Anchoring into the pruning metric.


\section{Method}

We introduce V-Skip, a method designed to compress multimodal CoT sequences. Departing from standard text-centric pruning, we theoretically reformulate the compression task as a VA-IB optimization. This formulation guides a dual-path selection mechanism that distinguishes between linguistic redundancy and visual necessity. The overall pipeline is illustrated in Figure~\ref{figure2}.

\subsection{Preliminaries and Problem Formulation}

We formalize the reasoning process of a MLLM as a probabilistic generation task. Let $\mathcal{I} = (\mathcal{V}, \mathcal{Q})$ denote the multimodal input context, comprising the visual input $\mathcal{V}$ and textual query $\mathcal{Q}$. The model generates a reasoning chain $\mathcal{C} = \{c_1, \dots, c_T\}$, where $c_{t}$ represents the $t$-th reasoning token in a sequence of total length $T$. The joint distribution is factorized autoregressively as:
\begin{equation}
\label{eq:joint_dist}
P_{\theta}(\mathcal{A}, \mathcal{C} | \mathcal{I}) = P_{\theta}(\mathcal{A} | \mathcal{C}, \mathcal{I}) \! \prod_{t=1}^{T} \! P_{\theta}(c_{t} | \mathcal{C}_{<t}, \mathcal{I}),
\end{equation}
where $P_{\theta}(\cdot)$ denotes the conditional probability distribution parameterized by the MLLM with weights $\theta$. The objective of token compression is to identify a sparse subsequence $\hat{\mathcal{C}} \subset \mathcal{C}$ with length $|\hat{\mathcal{C}}|\ll |\mathcal{C}|$ that maximizes the predictive likelihood of the answer $\mathcal{A}$ while minimizing inference latency. 

Standard text pruning methods typically operate by minimizing the negative log-likelihood of the sequence based on linguistic priors. However, directly applying this unimodal metric to MLLMs precipitates a critical failure mode we term Visual Amnesia.
In a multimodal context, visually grounded tokens often exhibit high linguistic perplexity because they cannot be inferred from the preceding text alone. As a result, probability-based pruners misidentify these high-entropy visual anchors as redundancy. This results in compressed chains that are fluent but factually ungrounded.

\subsection{Visual-Anchored Information Bottleneck}

To mitigate Visual Amnesia, we reformulate the compression task through the lens of the Information Bottleneck (IB) principle~\cite{tishby2000information,alemi2016deep}. Building upon this principle, we posit that a valid compressed rationale $\hat{\mathcal{C}}$ must satisfy two distinct information-theoretic criteria. The first is sufficiency, which ensures the retention of adequate semantic content to predict the answer $\mathcal{A}$. The second is grounding, which necessitates maintaining high mutual dependence with the visual input $\mathcal{V}$ to mitigate hallucination risks.

To operationalize these criteria, we propose the VA-IB objective. Unlike standard text compression which minimizes length based solely on linguistic priors, VA-IB explicitly imposes a visual constraint. We formulate the optimization as maximizing the joint utility subject to a length budget:
\begin{equation}
\label{eq:va-ib}
\begin{aligned}
\max_{\hat{\mathcal{C}}} \Big[ \underbrace{I(\hat{\mathcal{C}}; \mathcal{A})}_{\text{Sufficiency}} &+ \lambda \underbrace{I(\hat{\mathcal{C}}; \mathcal{V} \mid \mathcal{Q})}_{\text{Anchoring}} \Big] \\
s.t. \quad &|\hat{\mathcal{C}}| \le \gamma |\mathcal{C}|,
\end{aligned}
\end{equation}
where $\lambda$ controls the strength of visual grounding and $\gamma$ is the target compression ratio. The core lies in the anchoring term $I(\hat{\mathcal{C}}; \mathcal{V} \mid \mathcal{Q})$. To elucidate its physical meaning, we decompose this conditional mutual information using Shannon entropy:

\begin{equation}
\label{eq:mi-decomp}
I(\hat{\mathcal{C}}; \mathcal{V} \mid \mathcal{Q}) = H(\hat{\mathcal{C}} \mid \mathcal{Q}) - H(\hat{\mathcal{C}} \mid \mathcal{V}, \mathcal{Q}),
\end{equation}
where $H(\hat{\mathcal{C}}\mid\mathcal{Q})$ represents Textual Uncertainty and $H(\hat{\mathcal{C}} \mid \mathcal{V}, \mathcal{Q})$ measures Multimodal Uncertainty.
Maximizing Eq.~(\ref{eq:mi-decomp}) effectively selects tokens that are highly unpredictable from text alone but become deterministic given the image.

\subsection{Visual Anchoring Score via Attention}

\begin{figure*}[t]
  \centering
  \includegraphics[width=1.0\linewidth]{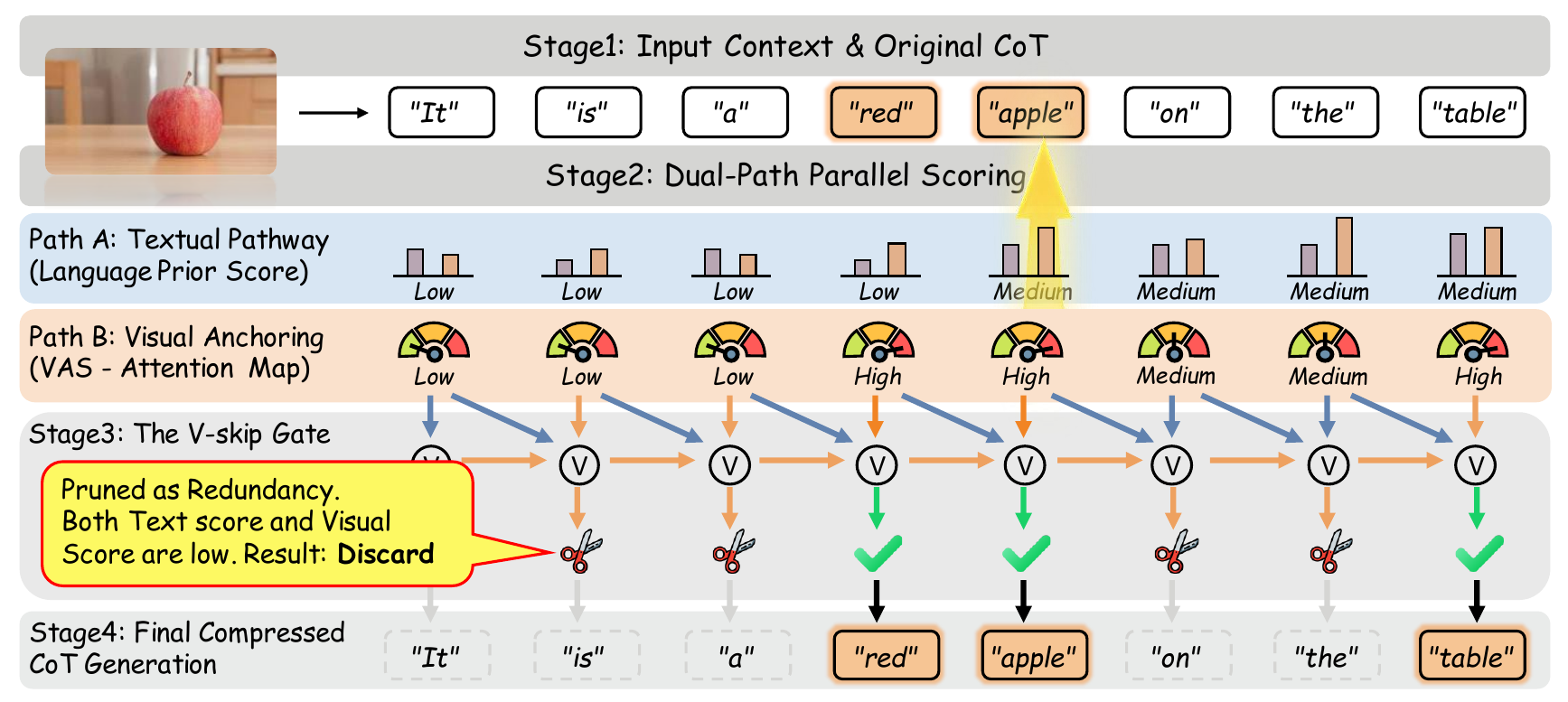}
  \caption{The Visual-Anchored Token Pruning Mechanism. The pruning process evaluates each generated token through a dual-pathway scoring framework. The Textual Pathway quantifies the linguistic necessity of a token by identifying redundant functional fillers, such as ``it'' or ``is'' based on their language prior scores. The Visual Pathway computes the VAS by aggregating cross-modal attention weights, which quantifies the grounding strength of a token relative to specific image regions. Finally, the V-Skip Gate applies a union-of-saliency operation to determine token retention. A token is pruned only if it is diagnosed as both linguistically redundant and visually irrelevant.}
  \label{figure3}
\end{figure*}

Optimizing the VA-IB objective requires evaluating the mutual information $I(c_t; \mathcal{V} \mid \mathcal{Q})$ for each token. Since direct computation over high-dimensional latent spaces is intractable, we derive an efficient proxy by leveraging the model's internal attention mechanism, detailed in Figure~\ref{figure3}.

In MLLMs trained with contrastive objectives, cross-modal attention weights serve as a variational lower bound for the pointwise mutual information between textual and visual representations. Specifically, the attention weight from a text token $c_t$ to a visual patch $v$ is proportional to their density ratio~\cite{oord2018representation}. This theoretical alignment implies that the cumulative attention mass directed toward the image manifests the degree of cross-modal dependence. We interpret the total attention flow as an empirical measure of a token’s grounding necessity. Based on this insight, we instantiate a dual-path scoring mechanism that evaluates linguistic redundancy and visual necessity.

\textbf{Textual Path $S_{\text{text}}$}\quad
To evaluate the linguistic redundancy of a generated token $c_t$, we quantify its information content using self-information. Formally, we define the textual score as the negative log-likelihood of the token conditioned on the multimodal context:
\begin{equation}
S_{\text{text}}(c_t) = -\log P_\theta(c_t \mid c_{<t}, \mathcal{Q}, \mathcal{V}),
\end{equation}
where $P_\theta(\cdot)$ is consistent with the formulation in Eq.~(\ref{eq:joint_dist}). Intuitively, tokens with high predictability ($e.g.,$ functional words like is or the) yield high probabilities and consequently low $S_{\text{text}}$ scores. These tokens provide minimal information gain and are considered candidates for pruning, provided they do not serve as visual anchors.

\textbf{Visual Path $S_{\text{vis}}$}\quad
Simultaneously, we measure the visual grounding by analyzing the attention flow from the generated text back to the visual context. Let $A^{(\ell, h)} \in \mathbb{R}^{T \times T}$ denote the attention weight matrix for layer $\ell$ and head $h$. Specifically, $A_{t, k}^{(\ell, h)}$ represents the attention weight from the query at position $t$ to the key at position $k$.
We define the total \textit{visual attention mass} $\Omega_{t}^{(\ell, h)}$ as the sum of attention weights directed towards the set of visual patch indices $\mathcal{I}_{\text{img}}$:
\begin{equation}
\Omega_{t}^{(\ell, h)} = \sum_{k \in \mathcal{I}_{\text{img}}} A_{t, k}^{(\ell, h)}.
\end{equation}

A high value of $\Omega_{t}$ indicates that the generation at step $t$ is heavily dependent on visual features, marking the corresponding token as a potential visual anchor.
However, recent mechanistic interpretability studies suggest that visual-text semantic alignment is not uniform. It typically peaks in the middle layers and is handled by sparse~\cite{ethayarajh2019contextual}, specialized induction heads rather than being distributed evenly~\cite{olsson2022context}. Simple averaging would dilute this signal with noise from syntactic heads. Therefore, we define the final Visual Anchoring Score (VAS) for the $t$-th token by aggregating $\Omega_{t}^{(\ell, h)}$ over a salient subset of layers $\mathcal{L}_{\text{focus}}$ and employing a max-pooling strategy over heads $\mathcal{H}$ to isolate the most confident grounding signal. This is formalized as:
\begin{equation}
\label{eq:vas}
S_{\text{vis}}(t) = \frac{1}{|\mathcal{L}_{\text{focus}}|} \sum_{\ell \in \mathcal{L}_{\text{focus}}} \max_{h \in \mathcal{H}} \left( \Omega_{t}^{(\ell, h)} \right).
\end{equation}

By defining $S_{\text{vis}}(t)$ in this manner, we effectively capture the Multimodal Uncertainty reduction described in Eq.~(\ref{eq:mi-decomp}). This ensures that tokens acting as visual anchors are identified and preserved regardless of their linguistic probability.

\subsection{The V-Skip Gating Mechanism}
\label{sec:gate}

We integrate the established dual utility metrics into the V-Skip Gate to construct a binary retention mask $M = \{m_1, \dots, m_T\}$, where $m_t \in \{0, 1\}$, identifying the optimal subsequence $\hat{\mathcal{C}}$ that satisfies the VA-IB objective defined in Eq.(\ref{eq:va-ib}). Unlike conventional pruning methods that rely on a single confidence threshold, V-Skip employs a Union-of-Saliency strategy.

This design is motivated by the complementary nature of our dual-path scores. $S_{\text{text}}$ identifies tokens essential for linguistic coherence, while $S_{\text{vis}}$ identifies tokens essential for visual grounding. Consequently, a token should be retained if it is salient in either modality.
Formally, given a target compression rate $\gamma$, we determine two dynamic thresholds $\tau_{\text{text}}$ and $\tau_{\text{vis}}$, corresponding to the $(1-\gamma)$-th percentile of their respective score distributions over the current sequence. The retention mask $m_t$ is determined by the disjunction of two boolean conditions, formulated as
\begin{equation}
\label{eq:vskip_logic}
m_t = \mathbb{I}\left(S_{\text{text}} \ge \tau_{\text{text}}\right) \lor \mathbb{I}\left(S_{\text{vis}} \ge \tau_{\text{vis}}\right),
\end{equation}
where $\mathbb{I}(\cdot)$ is the indicator function. The final compressed chain is obtained by filtering the sequence with this mask. By allowing visual saliency to override linguistic redundancy, this mechanism explicitly prevents the pruning of tokens that are linguistically predictable (low $S_{\text{text}}$) but visually essential (high $S_{\text{vis}}$). This ensures that critical visual descriptors, which are often treated as noise by unimodal compressors, are preserved, thereby maintaining the factual integrity of the reasoning chain.

\subsection{Efficient Inference via V-Skip Distillation}
\label{sec:distillation}

To eliminate the runtime overhead of computing dual-path scores, we distill the V-Skip policy into the model parameters via efficient instruction tuning. We first construct a distilled dataset $\mathcal{D}_{\text{distill}}$ offline by applying the retention mask (Eq.~(\ref{eq:vskip_logic})) to raw reasoning chains. Subsequently, we fine-tune the model using LoRA~\cite{hu2022lora} under standard autoregressive supervision. This optimization allows the model to internalize the pruning logic, enabling it to generate concise and visually anchored rationales without explicit attention analysis during inference.

\section{Experiments}
\label{sec:experiments}

\begin{table*}[t]
\centering

\caption{
Main results of Qwen2-VL-7B-Instruct on MMMU and DocVQA datasets. We report Accuracy (\%) for MMMU and ANLS (\%) for DocVQA. To evaluate efficiency, we detail the average number of generated reasoning tokens (\textbf{Tokens}), end-to-end inference latency (\textbf{Latency}), and the actual compression ratio (\textbf{ActRatio}). The symbol $\gamma$ denotes the target compression ratio set for the pruning algorithms. \textcolor{red}{$\downarrow$} indicates the performance drop compared to the original full model.}
\label{tab:main_results}
\resizebox{\textwidth}{!}{%
\renewcommand{\arraystretch}{0.95}
\begin{tabular}{l c cccc cccc}
\toprule
\multirow{2}{*}{\textbf{Methods}} & \multirow{2}{*}{\textbf{Ratio $\gamma$}} & \multicolumn{4}{c}{\textbf{MMMU}} & \multicolumn{4}{c}{\textbf{DocVQA}} \\
\cmidrule(lr){3-6} \cmidrule(lr){7-10}
 & & \textbf{Acc} & \textbf{Tokens} & \textbf{Latency} (s) & \textbf{ActRatio} & \textbf{ANLS} & \textbf{Tokens} & \textbf{Latency} (s) & \textbf{ActRatio} \\
\midrule
Original (Full) & - & 54.1 & 245.0 & 6.42 & - & 91.6 & 189.0 & 4.87 & - \\
\midrule
\multirow{3}{*}{Truncation} 
 & 0.9 & 50.8\scriptsize(\textcolor{red}{3.3$\downarrow$}) & 220.5 & 5.84 & 0.90 & 84.2\scriptsize(\textcolor{red}{7.4$\downarrow$}) & 170.1 & 4.52 & 0.90 \\
 & 0.7 & 44.5\scriptsize(\textcolor{red}{9.6$\downarrow$}) & 171.5 & 4.51 & 0.70 & 71.5\scriptsize(\textcolor{red}{20.1$\downarrow$}) & 132.3 & 3.51 & 0.70 \\
 & 0.5 & 38.5\scriptsize(\textcolor{red}{15.6$\downarrow$}) & 122.5 & 3.23 & 0.50 & 62.5\scriptsize(\textcolor{red}{29.1$\downarrow$}) & 94.5 & 2.57 & 0.50 \\
\midrule
\multirow{3}{*}{LLMLingua-2} 
 & 0.9 & 49.5\scriptsize(\textcolor{red}{4.6$\downarrow$}) & 223.2 & 6.03 & 0.91 & 78.4\scriptsize(\textcolor{red}{13.2$\downarrow$}) & 173.8 & 4.74 & 0.92 \\
 & 0.7 & 40.2\scriptsize(\textcolor{red}{13.9$\downarrow$}) & 166.6 & 4.97 & 0.68 & 55.6\scriptsize(\textcolor{red}{36.0$\downarrow$}) & 130.4 & 3.81 & 0.69 \\
 & 0.5 & 32.4\scriptsize(\textcolor{red}{21.7$\downarrow$}) & 115.1 & 3.73 & 0.47 & 38.5\scriptsize(\textcolor{red}{53.1$\downarrow$}) & 88.8 & 2.93 & 0.47 \\
\midrule
\multirow{3}{*}{ASCoT} 
 & 0.9 & 50.1\scriptsize(\textcolor{red}{4.0$\downarrow$}) & 218.0 & 6.12 & 0.89 & 79.8\scriptsize(\textcolor{red}{11.8$\downarrow$}) & 168.2 & 4.78 & 0.89 \\
 & 0.7 & 41.8\scriptsize(\textcolor{red}{12.3$\downarrow$}) & 176.4 & 5.20 & 0.72 & 58.2\scriptsize(\textcolor{red}{33.4$\downarrow$}) & 136.1 & 4.03 & 0.72 \\
 & 0.5 & 33.1\scriptsize(\textcolor{red}{21.0$\downarrow$}) & 124.9 & 3.91 & 0.51 & 40.2\scriptsize(\textcolor{red}{51.4$\downarrow$}) & 90.7 & 3.10 & 0.48 \\
\midrule
\multirow{6}{*}{V-Skip (Ours)} 
 & 1.0 & 54.1\scriptsize(\textcolor{red}{0.0$\downarrow$}) & 245.0 & 6.54 & 1.00 & 91.6\scriptsize(\textcolor{red}{0.0$\downarrow$}) & 189.0 & 5.09 & 1.00 \\
 & 0.9 & 53.6\scriptsize(\textcolor{red}{0.5$\downarrow$}) & 227.8 & 5.89 & 0.93 & 90.8\scriptsize(\textcolor{red}{0.8$\downarrow$}) & 172.1 & 4.61 & 0.91 \\
 & 0.8 & 52.8\scriptsize(\textcolor{red}{1.3$\downarrow$}) & 193.6 & 5.24 & 0.79 & 89.5\scriptsize(\textcolor{red}{2.1$\downarrow$}) & 156.9 & 4.08 & 0.83 \\
 & 0.7 & 51.5\scriptsize(\textcolor{red}{2.6$\downarrow$}) & 173.4 & 4.65 & 0.71 & 87.9\scriptsize(\textcolor{red}{3.7$\downarrow$}) & 132.4 & 3.68 & 0.70 \\
 & 0.6 & 50.1\scriptsize(\textcolor{red}{4.0$\downarrow$}) & 151.9 & 4.05 & 0.62 & 85.8\scriptsize(\textcolor{red}{5.8$\downarrow$}) & 115.3 & 3.19 & 0.61 \\
 & 0.5 & 48.2\scriptsize(\textcolor{red}{5.9$\downarrow$}) & 120.1 & 3.49 & 0.49 & 83.7\scriptsize(\textcolor{red}{7.9$\downarrow$}) & 98.4 & 2.71 & 0.52 \\
\bottomrule
\end{tabular}
}
\end{table*}

\subsection{Experimental Setup}
\label{sec:setup}
\textbf{Models and Architectures}\quad 
To evaluate V-Skip across varying scales and architectures, we employ the Qwen2-VL-Instruct series (encompassing the 2B, 7B, and 72B variants)~\cite{wang2024qwen2} alongside Llama-3.2-11B-Vision-Instruct~\cite{dubey2024llama}. We implement V-Skip by applying LoRA to the attention layers while freezing the pre-trained visual and language backbones. All experiments are conducted on a server equipped with 8 NVIDIA RTX 3090 GPUs.

\textbf{Datasets and Metrics}\quad
To evaluate the efficacy of V-Skip, we employ two distinct benchmarks. We utilize the MMMU dataset to assess complex multidisciplinary reasoning~\cite{yue2024mmmu}. Furthermore, we leverage the DocVQA benchmark to strictly evaluate fine-grained OCR and spatial grounding abilities~\cite{mathew2021docvqa}. We report Accuracy, End-to-End Latency, and the Actual Compression Ratio (ActRatio). Detailed experimental configurations and dataset splits are provided in the Appendix \ref{sec:appendix_implementation}.

\textbf{Baselines}\quad To strictly evaluate the contribution of our visual-anchored mechanism, we compare V-Skip against both text-centric and multimodal compression strategies. For text-centric methods, we employ ASCoT~\cite{zhang2025ascot} and LLMLingua-2~\cite{pan2024llmlingua}, which rely solely on linguistic probability distributions to identify redundancy. For multimodal comparison, we implement a Visual-Attention Pruning as a lower bound for performance stability~\cite{wang2024qwen2}.

\subsection{Main Results}
\label{sec:main_results}

The quantitative comparison on Qwen2-VL-7B detailed in Table~\ref{tab:main_results}. Text-centric baselines suffer significant performance degradation on the fine-grained DocVQA task. Notably, LLMLingua-2 drops over 53\% at $\gamma=0.5$. In contrast, V-Skip effectively preserves visual anchors, retaining 83.7\% ANLS and outperforming LLMLingua-2 by 45.2\%. Similarly, on MMMU, V-Skip limits accuracy loss to 5.9\%, whereas baselines degrade by over 20\%. In terms of efficiency, V-Skip achieves a $1.8\times$ speedup on DocVQA (2.71s). Crucially, it processes faster than LLMLingua-2 (2.93s) despite generating longer sequences, confirming the runtime advantage of our zero-overhead LoRA distillation over calculation-heavy online filtering.

\begin{figure}[t]
  \centering
  \includegraphics[width=1.0\linewidth]{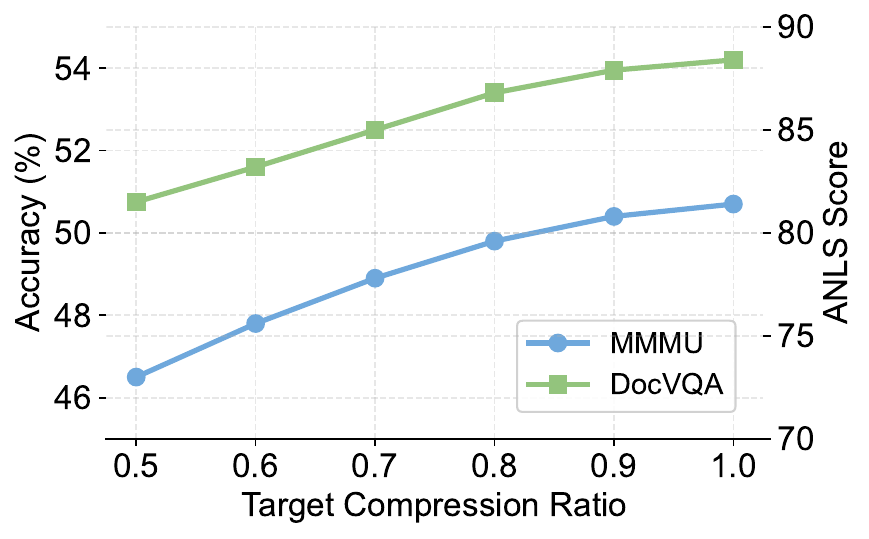}
  \caption{Performance robustness of V-Skip on Llama-3.2-11B-Vision-Instruct across varying compression ratios. The dual-axis plot displays the impact of the target compression ratio $\gamma$ (x-axis) on model performance. The left axis (blue circle) reports the Accuracy (\%) for the MMMU, while the right axis (green square) indicates the ANLS Score for the DocVQA.}
  \label{figure4}
\end{figure}

To verify the scalability of our method, we extend our evaluation to the Llama-3.2-11B-Vision-Instruct model, as illustrated in Figure~\ref{figure4}. The V-Skip method maintains a flat performance curve across varying compression ratios for both MMMU and DocVQA tasks. Specifically, even under aggressive pruning ($\gamma=0.5$), Llama-3.2 with V-Skip retains over 91\% of its original performance on both benchmarks. This demonstrates that the principle of Visual Anchoring is model-agnostic and essential for robust multimodal compression.

\textbf{Robustness Increases with Model Size}\quad We further investigate the scalability of V-Skip across the Qwen2-VL family (2B, 7B, and Large) on the complex MMMU benchmark. As detailed in Table \ref{tab:scaling}, we observe a distinct Positive Scaling Law regarding pruning robustness. The lightweight 2B model exhibits high sensitivity to token removal, suffering a significant performance degradation of 8.5\%. We attribute this to the fragility of reasoning chains in parameter constrained models, where information is densely packed and lacks redundancy. Removing intermediate steps in such concise chains often leads to logical fractures.
Conversely, the 72B model demonstrates superior resilience. Despite removing 50\% of the tokens, the accuracy drop is limited to just 3.2\%, effectively halving the loss observed in the 7B baseline. This result suggests that as model capacity increases, the generated CoT becomes inherently more verbose and redundant.

\subsection{Analysis}
\label{sec:analysis}

We conduct an in-depth analysis to interpret the underlying mechanisms of V-Skip. We first quantify the preservation of fine-grained descriptors through a visual attribute retention analysis. Complementing this quantitative evidence, we examine the information entropy mismatch through a qualitative case study. Ultimately, we link these granular behaviors to broader model safety, evaluating how preserving visual grounding acts as a stabilizer to mitigate hallucination bias.

\subsubsection{Visual Attribute Retention Analysis}
To investigate the underlying cause of the performance degradation observed in text-only baselines, we hypothesize that text-centric pruning inadvertently discards visual descriptors ($e.g.,$ colors, shapes) due to their low linguistic perplexity. To quantify this, we measure the Visual Attribute Retention Rate (VARR) on MMMU using the GQA vocabulary. As shown in Table~\ref{tab:varr_breakdown}, probability-based baselines suffer significant information loss. Notably, LLMLingua-2 retains only 42.5\% of color tokens, treating the majority as redundancy. In contrast, V-Skip effectively rescues these anchors via visual saliency. At the same compression ratio ($\gamma=0.5$), V-Skip achieves 89.4\% retention for colors (+46.9\% over LLMLingua-2) and 91.2\% for objects. This empirical evidence confirms that V-Skip successfully anchors the compressed CoT to the visual modality, preventing the reasoning collapse observed in baselines.

\begin{table}[t]
\centering
\small
\caption{Scaling analysis on the MMMU. We evaluate the impact of V-Skip across the Qwen2-VL series models (2B, 7B, and 72B) with a fixed target compression ratio of $\gamma=0.5$. We reports the accuracy of the uncompressed baseline (Full CoT), the compressed model (V-Skip), and the absolute performance decrease ($\Delta$).}
\label{tab:scaling}
\setlength{\tabcolsep}{10pt}
\begin{tabular}{l|ccc} 
\toprule
\textbf{Model Size} & \textbf{Full CoT} & \textbf{V-Skip} & \textbf{Drop ($\Delta$)} \\
\midrule
2B (Small)   & 41.2 & 32.7 & \textcolor{red}{$-$8.5} \\
7B (Base)    & 54.1 & 48.2 & \textcolor{red}{$-$5.9} \\
72B (Large)  & 64.5 & 61.3 & \textcolor{red}{$-$3.2} \\
\bottomrule
\end{tabular}
\end{table}

\subsubsection{Case Study}
\label{sec:qualitative_main}

To empirically validate the mechanism behind the performance divergence observed in Table~\ref{tab:main_results}, we present a qualitative case study in Figure~\ref{figure5}. The task requires the model to extract a specific invoice total \$45.20.
This case exposes a fundamental Information Entropy Mismatch between modalities. From a unimodal linguistic perspective, specific numerical values often appear as high entropy outliers, they lack semantic redundancy with the preceding context. Consequently, probability based pruners like LLMLingua-2 misinterpret these critical tokens as noise, leading to the erroneous pruning of the answer itself.

V-Skip resolves this conflict by integrating the Visual Score $S_{\text{vis}}$. Although the token ``\$45.20'' exhibits low linguistic coherence, it triggers a massive activation in the cross-attention layers, indicating a strong dependency on image pixels. By detecting this signal, V-Skip overrides the low textual probability and identifies the token as an indispensable bridge to the visual context.
This enables V-Skip to function as a grounding-aware filter, systematically preserving tokens characterized by high visual saliency despite their high linguistic perplexity, which are essential for precise visual reasoning.

\begin{table}[t]
\centering
\small
\caption{Visual Attribute Retention Rate (VARR) analysis on the MMMU dataset. We categorize tokens within the CoT into specific visual attribute classes (Color, Shape, Object) using the GQA vocabulary. The table reports the percentage of retained tokens for each category at a fixed compression ratio of $\gamma=0.5$.}
\label{tab:varr_breakdown}
\setlength{\tabcolsep}{8pt}
\begin{tabular}{l|ccc}
\toprule
\textbf{Method} & \textbf{Color} & \textbf{Shape} & \textbf{Object} \\
\midrule
LLMLingua-2 & 42.5\% & 55.1\% & 64.8\% \\
ASCoT & 48.2\% & 58.7\% & 69.5\% \\
\midrule
V-Skip (Ours) & \textbf{89.4\%} & \textbf{86.3\%} & \textbf{91.2\%} \\
\bottomrule
\end{tabular}
\end{table}

\subsubsection{Mitigating Hallucination Bias}
A critical concern in MLLMs is the tendency to hallucinate non existent objects. To determine whether token pruning exacerbates this issue,
we investigate whether pruning exacerbates object hallucination using the balanced POPE benchmark. As detailed in Table~\ref{tab:pope_breakdown}, text-centric baselines exhibit a severe Yes-Bias, with positive response rates surging to 66.8\% (ASCoT) and 64.5\% (LLMLingua-2). While this inflates Recall, it severely degrades Precision to roughly 63.5\%, indicating a reliance on parametric priors when visual anchors are stripped. In contrast, V-Skip effectively preserves visual grounding via $S_{\text{vis}}$, maintaining a neutral Yes-Ratio of 51.2\% that closely mirrors the uncompressed baseline (50.4\%). Consequently, V-Skip achieves a balanced F1 score of 88.9, demonstrating its capability to compress reasoning chains without compromising factual integrity.

\begin{table}[t]
\centering
\small
\caption{
Hallucination evaluation on the POPE benchmark. We report standard classification metrics (Accuracy, Precision, Recall, F1) and the Yes-Ratio (the proportion of positive responses generated by the model).}
\label{tab:pope_breakdown}
\setlength{\tabcolsep}{5pt}
\begin{tabular}{l|c|cccc}
\toprule
\textbf{Method} & \textbf{Yes-Ratio} & \textbf{Acc} & \textbf{Prec} & \textbf{Recall} & \textbf{F1} \\
\midrule
Full CoT & 50.4\% & 89.2 & 89.8 & 88.5 & 89.1 \\
\midrule
ASCoT & 66.8\% & 73.1 & 63.5 & \textbf{98.2} & 77.1 \\
LLMLingua-2 & 64.5\% & 75.4 & 66.8 & 96.5 & 78.9 \\
\midrule
V-Skip (Ours) & \textbf{51.2\%} & \textbf{88.9} & \textbf{89.1} & 88.7 & \textbf{88.9} \\
\bottomrule
\end{tabular}
\end{table}

\subsection{Ablation Experiments}
\label{sec:ablation}

To validate our dual-path design, we decouple the gating components on the MMMU dataset. As shown in Table~\ref{tab:ablation_mechanism}, results reveal a distinct trade-off between linguistic coherence and visual grounding. Relying solely on textual probability (Row B) minimizes syntax errors (1.1\%) but sacrifices visual recall (low VARR). Conversely, the Vision-Only method (Row C) maximizes VARR (92.4\%) but disrupts syntactic fluency by pruning essential connectives. V-Skip's Union strategy (Row E) effectively resolves this conflict, achieving the highest accuracy (48.2\%) by synergizing linguistic structure with visual anchors. Notably, the Intersection strategy (Row D) underperforms (30.1\%) due to overly restrictive filtering, which significantly reduces information recall. Additional sensitivity analyses are provided in Appendix~\ref{sec:appendix_ablation}.

\begin{table}[t]
\caption{
Component analysis of the V-Skip gating mechanism on the MMMU dataset. We evaluate the impact of individual scoring components ($S_{\text{text}}$, $S_{\text{vis}}$) and their combinations on model performance. The table reports classification Accuracy, VARR, and Syntax Error rate (Err) at a fixed target compression ratio of $\gamma=0.5$.}
\label{tab:ablation_mechanism}
\centering
\small
\setlength{\tabcolsep}{4pt}
\begin{tabular}{l|c|c|c}
\toprule
\textbf{Pruning Strategy} & \textbf{Acc (\%)} & \textbf{VARR (\%)} $\uparrow$ & \textbf{Err (\%)} $\downarrow$ \\
\midrule
(A) Random & 27.5 & 49.8 & 14.5 \\
(B) Text-Only & 32.4 & 58.8 & \textbf{1.1} \\
(C) Vision-Only & 40.5 & \textbf{92.4} & 11.2 \\
(D) Intersection & 30.1 & 52.4 & 3.5 \\
\midrule
(E) V-Skip Union & \textbf{48.2} & 89.0 & 1.3 \\
\bottomrule
\end{tabular}
\end{table}

\section{Conclusion}
In this work, we address the efficiency bottleneck of MLLMs by identifying Visual Amnesia where text-centric pruning inadvertently discards tokens essential for visual grounding. To resolve this, we propose V-Skip, a method theoretically grounded in the VA-IB principle. By synergizing linguistic surprisal with cross-modal attention flow and distilling this policy into a lightweight LoRA adapter, V-Skip achieves a $2.9\times$ inference speedup with negligible performance loss. Crucially, empirical results demonstrate that V-Skip outperforms state-of-the-art baselines by over 45\% on fine-grained visual tasks while significantly mitigating object hallucination. 
Although V-Skip is primarily validated within the context of textual CoT, we hope this work serves as a stepping stone for future research to explore similar alignment based compression strategies in broader modalities, such as dynamic video streams or audio visual interactions.

\section*{Limitations}
Despite V-Skip's robustness, we acknowledge two limitations. First, it depends on the base model's cross-modal alignment, potentially degrading on architectures with attention. Second, unlike training-free methods, it necessitates offline distillation, incurring additional training overhead.




\bibliography{custom}
\newpage
\appendix

\section{More details of our Algorithms}

\subsection{Algorithms}
\label{sec:appendix_algo}
We summarize the complete V-Skip pipeline in Algorithm \ref{alg:vskip_framework}. The process consists of two sequential phases. For phase 1, we first utilize a frozen Teacher MLLM to generate full CoT rationales. For every token in the sequence, we compute two utility metrics, the textual information score $S_{\text{text}}$ derived from linguistic probability, and the visual anchoring score $S_{\text{vis}}$ derived from cross-modal attention maps (as defined in Eq.~(\ref{eq:vas})). An Union-of-Saliency gate is then applied. A token is retained if it exceeds the dynamic threshold in either the textual or visual modality. This mechanism explicitly prevents Visual Amnesia by rescuing high entropy visual tokens that would be pruned by text only metrics. For phase 2, to avoid the computational overhead of calculating attention scores at runtime, we distill the pruning policy into a student model. Using the compressed rationales $\hat{\mathcal{C}}$ generated in Phase 1 as targets, we fine-tune the Base Student Model via LoRA. This allows the student to internalize the V-Skip logic, enabling it to directly generate concise, visually grounded rationales during inference with zero additional latency.

\begin{algorithm}[h]
\caption{The V-Skip Framework: From Dual-Path Pruning to Efficient Distillation}
\label{alg:vskip_framework}
\begin{algorithmic}[1]
\Require Dataset $\mathcal{D}$, Teacher MLLM $\theta_{\text{teach}}$, Base Student $\theta_{\text{student}}$, Target Rate $\gamma$
\Ensure Efficient Student Model $\theta_{\text{student}}$

\State \textbf{// Phase 1: Data Construction via Dual-Path Pruning}
\State Initialize training set $\mathcal{D}_{\text{train}} \leftarrow \emptyset$

\For{each sample $(\mathcal{V}, \mathcal{Q}) \in \mathcal{D}$}
    \State $\mathcal{C} \leftarrow \text{GenerateCoT}(\theta_{\text{teach}}, \mathcal{V}, \mathcal{Q})$
    
    \State \textbf{// Calculate Scores per Token $c_t$}
    \State $S_{\text{text}} \leftarrow -\log P(c_t \mid c_{<t})$ 
    \State $S_{\text{vis}} \leftarrow \text{AggAttention}(c_t \to \mathcal{V})$ 
    
    \State \textbf{// Dynamic Thresholding}
    \State $\tau_{\text{text}}, \tau_{\text{vis}}\leftarrow \text{Perc}((1-\gamma), \{S_{\text{text}}\}, \{S_{\text{vis}}\})$
    
    \State \textbf{// Apply Union Gate}
    \State $\hat{\mathcal{C}} \leftarrow \emptyset$
    \For{$t = 1$ \textbf{to} Length($\mathcal{C}$)}
        \If{$S_{\text{text}}^{(t)} \ge \tau_{\text{text}} \lor S_{\text{vis}}^{(t)} \ge \tau_{\text{vis}}$}
            \State Append $c_t$ to $\hat{\mathcal{C}}$
        \EndIf
    \EndFor
    \State Add $(\mathcal{V}, \mathcal{Q}, \hat{\mathcal{C}})$ to $\mathcal{D}_{\text{train}}$
\EndFor

\State \textbf{// Phase 2: LoRA Fine-tuning}
\State Inject LoRA adapters into $\theta_{\text{student}}$
\While{not converged}
    \State Sample batch from $\mathcal{D}_{\text{train}}$
    \State $\mathcal{L} \leftarrow -\log P_{\theta_{\text{student}}}(\hat{\mathcal{C}} \mid \mathcal{V}, \mathcal{Q})$
    \State Update parameters via SGD
\EndWhile

\State \textbf{return} $\theta_{\text{student}}$
\end{algorithmic}
\end{algorithm}

\subsection{Qualitative Analysis: What is Skipped?}
\label{sec:qualitative}

\begin{figure*}[t]
  \centering
  \includegraphics[width=0.9\linewidth]{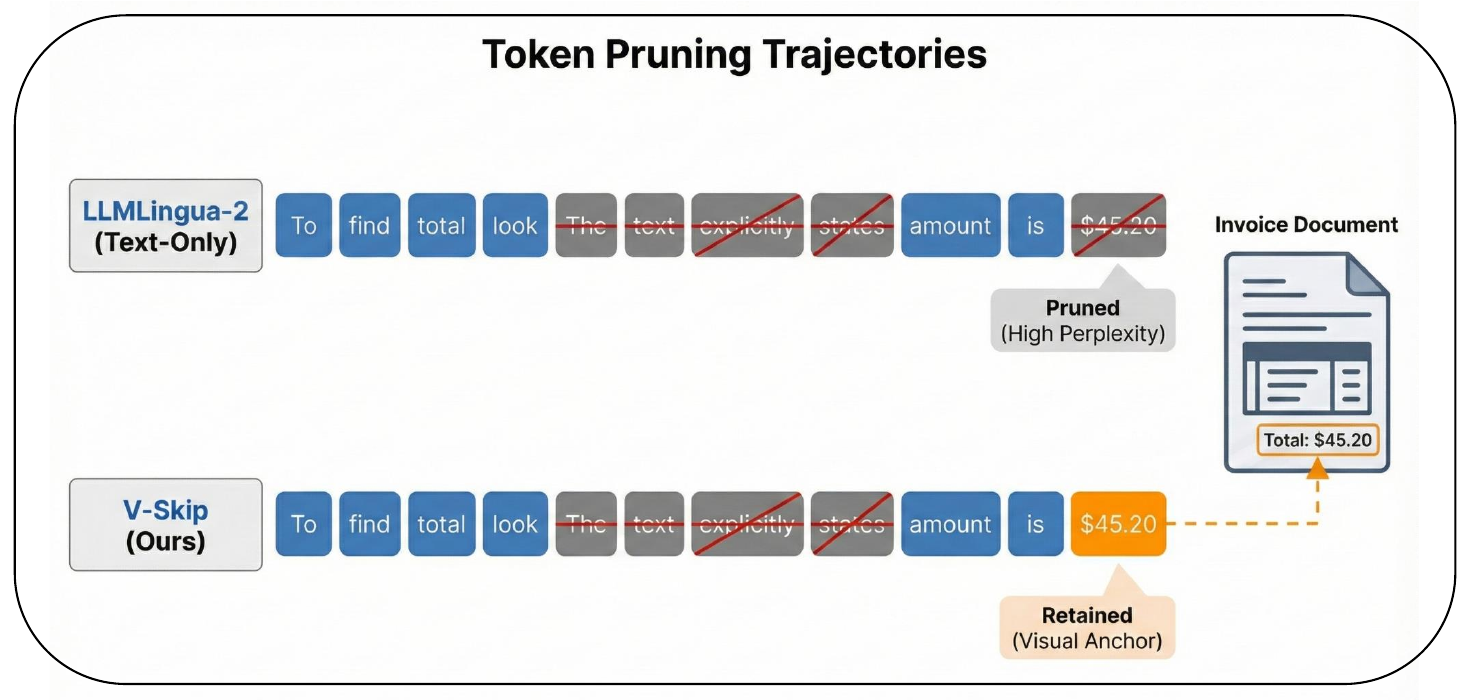}
  \caption{Qualitative comparison on DocVQA illustrating the Information Entropy Mismatch. V-Skip successfully retains the high-surprisal token \$45.20 by leveraging visual attention, whereas text-only pruning fails.}
  \label{figure5}
\end{figure*}

To provide an intuitive understanding of the pruning dynamics, Figure \ref{fig:case_study} visualizes the token-level compression trajectory for the DocVQA invoice case discussed in Section \ref{sec:qualitative_main}.

We compare the retention masks generated by LLMLingua-2 and V-Skip at a compression ratio of $\gamma=0.5$. Deleted tokens are marked with strikethroughs.

LLMLingua-2 successfully removes linguistic fillers, but disastrously removes the answer \$45.20 due to its high linguistic perplexity. This results in an incoherent statement.
V-Skip demonstrates the ability to learn semantic shortcuts. It skips the same linguistic fillers as the baseline but selectively retains the high entropy visual token \$45.20. The resulting compressed CoT (Find total look bottom right... total amount is \$45.20) retains the minimal sufficient statistics required for the final prediction.

This visualization confirms that V-Skip operates not merely as a text compressor, but as a grounding-aware filter that aligns latent reasoning with visual evidence.

\section{Implementation Details}
\label{sec:appendix_implementation}

\subsection{Datasets}
We first utilize the MMMU benchmark, a massive multi-discipline evaluation suite comprising collegiate-level tasks across disciplines such as Art, Science, and Engineering. Unlike simple recognition tasks, MMMU demands deep multimodal reasoning and expert-level knowledge application. We select this dataset to rigorously assess whether the compressed rationales generated by V-Skip maintain the logical coherence and semantic integrity required for complex problem solving.
Complementing this, we leverage the DocVQA benchmark, which consists of diverse document images ($e.g.,$ invoices, forms, reports) paired with questions requiring precise Optical Character Recognition (OCR) and spatial layout understanding. In DocVQA, correct answers often hinge on high entropy visual tokens that appear statistically random to language models.
To measure performance and controllability, we report Accuracy, End-to-End Latency, and the ActRatio.

\subsection{Detailed Evaluation Metrics}

To provide a comprehensive assessment of V-Skip's performance, we report three key efficiency metrics alongside task accuracy. Here, we provide the precise definitions and measurement protocols for each metric reported in Table~\ref{tab:main_results}.

\textbf{Tokens}
This metric quantifies the verbosity of the reasoning chain. For a given dataset $\mathcal{D}$, let $\hat{\mathcal{C}}_i$ denote the generated Chain-of-Thought sequence for the $i$-th sample. The Average Generated Tokens, $L_{\text{avg}}$, is defined as:
\begin{equation}
L_{\text{avg}} = \frac{1}{|\mathcal{D}|} \sum_{i=1}^{|\mathcal{D}|} \text{Length}(\hat{\mathcal{C}}_i),
\end{equation}
where $\text{Length}(\cdot)$ counts the number of tokens decoded by the model, excluding the input prompt. This metric serves as a proxy for the information density of the reasoning process. A significantly lower $L_{\text{avg}}$ with maintained accuracy indicates a more concise and efficient reasoning path.

\textbf{Latency}
We report the wall-clock inference time to strictly evaluate the real-world efficiency of our method. Unlike theoretical complexity measures (FLOPs), this metric captures the total computational cost, including method specific overheads.

\textbf{ActRatio}
While we set a target compression ratio $\gamma$ (e.g., 0.5 implies retaining 50\% of tokens) as a hyperparameter, the realized compression often varies due to the distinct pruning logic of each algorithm. We define the ActRatio as:
\begin{equation}
\text{ActRatio} = \frac{\sum \text{Length}(\hat{\mathcal{C}}_{\text{compressed}})}{\sum \text{Length}(\mathcal{C}_{\text{original}})}.
\end{equation}

The ActRatio is dynamic. Consequently, if an image requires extensive visual grounding, V-Skip may yield an ActRatio slightly higher than $\gamma$ to rescue critical visual tokens. Conversely, for simple images, it may compress more aggressively. We report this metric to transparently demonstrate the distinct behavior of our visual-anchored mechanism compared to rigid text-only truncation.

\textbf{Acc for MMMU}
The MMMU benchmark comprises multidisciplinary tasks requiring expert level reasoning, formatted primarily as multiple choice questions. We report the standard Top-1 Accuracy:
\begin{equation}
\text{Acc} = \frac{1}{N} \sum_{i=1}^{N} \mathbb{I}(\hat{y}_i = y_i),
\end{equation}
where $N$ is the total number of evaluation samples, $\hat{y}_i$ is the model's predicted option, $y_i$ is the ground truth, and $\mathbb{I}(\cdot)$ is the indicator function.

\textbf{ANLS for DocVQA}
For the DocVQA dataset, where answers are often extracted text spans susceptible to minor Optical Character Recognition (OCR) variations, exact match accuracy is overly penalizing. Following standard evaluation protocols for document intelligence, we employ the ANLS metric.

ANLS measures the edit distance between the predicted text sequence and the ground truth, normalized by the sequence length. For a prediction $P$ and ground truth $G$, the score $s(P, G)$ is defined as:
\begin{equation}
\begin{aligned}
& \text{Let } M = \max(|P|, |G|). \\
& s(P, G) = \begin{cases} 
1 - \frac{d_{L}(P, G)}{M}, & \text{if } d_{L}(P, G) < \tau \cdot M \\
0, & \text{otherwise}
\end{cases}
\end{aligned}
\end{equation}
where $d_{L}(\cdot)$ denotes the Levenshtein distance, $| \cdot |$ denotes the string length, and $\tau=0.5$ is the threshold penalty. The final ANLS reported in Table~\ref{tab:main_results} is the mean score over all samples, scaled to a percentage ($0-100$).

\subsection{Dataset Splits and Evaluation}
For all benchmarking experiments reported in the main text, we strictly utilize the official validation splits of both MMMU and DocVQA. This protocol ensures a consistent and fair comparison with baseline methods while preventing any potential test set leakage.

\subsection{Training Configurations}
We implement the V-Skip distillation process using LoRA to efficiently fine-tune the attention mechanisms of the pre-trained MLLMs. This allows the model to internalize the pruning policy with minimal computational overhead. We set the LoRA rank $r=16$ and the scaling factor $\alpha=32$. To fully capture the attention-based pruning logic, we apply the low-rank adapters to all linear projection layers within the self-attention modules ($W_q, W_k, W_v, W_o$). The LoRA dropout rate is set to 0.05 to prevent overfitting during the distillation phase.

The models are optimized using AdamW with a learning rate of $2 \times 10^{-4}$ and a cosine learning rate scheduler featuring a warm-up ratio of 0.03. We fine-tune the student models for a total of 3 epochs, which we found empirically sufficient for the model to learn the V-Skip policy without degrading its generation quality. We utilize a global batch size of 16 to ensure stable convergence on our 8$\times$RTX 3090 GPU infrastructure.

\begin{table*}[t]
\caption{Sensitivity to visual layer selection on the MMMU dataset. We evaluate the impact of deriving the Visual Anchoring Score $S_{\text{vis}}$ from different ranges of Transformer layers. The table reports the model accuracy and the performance deviation ($\Delta$) relative to the Middle Layers setting, conducted at a fixed compression ratio of $\gamma=0.5$.}
\label{tab:layer_ablation}
\centering
\small
\setlength{\tabcolsep}{10pt}
\begin{tabular}{l|c|cc}
\toprule
\textbf{Layer Selection} & \textbf{Depth Range} & \textbf{Acc (\%)} & \textbf{Drop ($\Delta$)} \\
\midrule
Early Layers & $0\% - 25\%$ & 41.5 & $-$6.7 \\
Late Layers & $75\% - 100\%$ & 46.8 & $-$1.4 \\
All Layers & $0\% - 100\%$ & 47.1 & $-$1.1 \\
\midrule
Middle (Ours) & \textbf{25\% - 75\%} & \textbf{48.2} & \textbf{$-$} \\
\bottomrule
\end{tabular}
\end{table*}

\begin{table*}[t]
\caption{
Impact of LoRA Rank ($r$) on performance and efficiency. We evaluate the sensitivity of the model to varying adapter sizes on the MMMU dataset. The table reports the proportion of trainable parameters, classification accuracy, and peak training VRAM usage at a fixed compression ratio of $\gamma=0.5$.}
\label{tab:rank_ablation}
\centering
\small
\setlength{\tabcolsep}{8pt}
\begin{tabular}{c|ccc}
\toprule
\textbf{LoRA Rank ($r$)} & \textbf{Trainable Params} & \textbf{Acc (\%)} & \textbf{Training VRAM} \\
\midrule
4 & 0.04\% & 46.5 & 18.2 GB \\
8 & 0.08\% & 47.6 & 18.3 GB \\
16 (Ours) & \textbf{0.16\%} & \textbf{48.2} & \textbf{18.5 GB} \\
32 & 0.32\% & 48.3 & 18.9 GB \\
64 & 0.64\% & 48.3 & 19.6 GB \\
\bottomrule
\end{tabular}
\end{table*}

\begin{figure*}[!t]
\centering
\begin{tcolorbox}[colback=blue!5!white,colframe=blue!75!black,title=\textbf{Qualitative Comparison on DocVQA}]
\small
\textbf{Question:} What is the invoice total shown at the bottom? \\
\textbf{Image Context:} An invoice image containing the text Total: \$45.20 in small print.

\vspace{0.1cm}
\hrule
\vspace{0.1cm}

\textbf{Original CoT:} \\
To find the total, look at the bottom right corner. \textcolor{gray}{The text explicitly states that} the total amount \textcolor{gray}{of the invoice} is \textbf{\$45.20}. \textcolor{gray}{Therefore, the answer is} \$45.20.

\vspace{0.1cm}
\textbf{LLMLingua-2 (Text-Only Pruning):} \\
To find total look bottom right corner. \sout{The text explicitly states that} total amount \sout{of the invoice} is \sout{\textbf{\$45.20}}. \sout{Therefore answer is} \sout{\$45.20}. \\
\textit{$\rightarrow$ Prediction: "Total amount is." (Incorrect)}

\vspace{0.1cm}
\textbf{V-Skip (Ours):} \\
Find total look bottom right. \sout{The text explicitly states that} total amount \sout{of the invoice} is \textbf{\$45.20}. \sout{Therefore answer} \$45.20. \\
\textit{$\rightarrow$ Prediction: "\$45.20" (Correct)}
\end{tcolorbox}
\vspace{0pt plus 1fill}
\caption{Qualitative Comparison. Black strikethrough text denotes pruned tokens. Note that LLMLingua-2 prunes the crucial value \$45.20 because numbers often have high perplexity in isolation. V-Skip retains it because the Visual Anchoring Score is high.}
\label{fig:case_study}
\end{figure*}

\section{Additional Ablation Studies}
\label{sec:appendix_ablation}

In this section, we provide supplementary experiments to further characterize the robustness and implementation choices of V-Skip.

\subsection{Sensitivity to Visual Layer Selection}
\label{sec:ablation_layer}

The calculation of the Visual Anchoring Score $S_{\text{vis}}$ relies on aggregating cross-attention weights. However, interpretability studies suggest that visual text alignment is stratified across the Transformer depth. To verify this hypothesis and identify the optimal signal source, we conduct a sensitivity analysis on the MMMU dataset, as summarized in Table~\ref{tab:layer_ablation}.

Relying on shallow layers results in the 41.5\% performance drop. These layers primarily encode low level visual features and basic syntactic information. Consequently, the attention maps here are too noisy to reflect high level semantic importance, leading to the erroneous pruning of critical reasoning tokens.
The final layers also yield suboptimal results (46.8\%). While deep layers capture rich semantics, they suffer from abstraction drift as the representation methods the output layer, it becomes highly specialized for the next token prediction objective, diluting the direct correspondence with specific image patches.
The intermediate layers serve as the optimal source, achieving the peak accuracy of 48.2\%. This confirms that the semantic bridging between visual embeddings and textual concepts is strongest in the middle of the network. By aggregating attention from this range, V-Skip captures the most robust alignment signals for visual grounding.

\subsection{Hyperparameter Robustness: LoRA Rank Analysis}
\label{sec:ablation_rank}

V-Skip utilizes LoRA to efficiently fine-tune the gating mechanism. To determine the optimal adapter size, we conduct a sensitivity analysis on the LoRA rank ($r$) ranging from 4 to 64, evaluating both classification accuracy on the MMMU dataset and computational efficiency (VRAM usage).

Table~\ref{tab:rank_ablation} illustrates the trade-off between model capacity and resource consumption. At $r=4$, the model exhibits a performance deficit, reaching only 46.5\% accuracy. This suggests that the extremely constrained parameter budget (0.04\%) creates a representational bottleneck, preventing the adapter from effectively learning the complex cross-modal alignment patterns necessary for precise pruning.
Increasing the rank beyond 16 yields diminishing returns. While scaling $r$ to 32 or 64 marginally improves accuracy to 48.3\% (+0.1\%), it comes at the cost of doubled parameter counts and increased memory overhead. Consequently, we identify $r=16$ as the Pareto-optimal configuration. It achieves near peak performance (48.2\%) while maintaining a minimal memory footprint (18.5 GB), demonstrating that V-Skip is highly parameter efficient and does not require extensive fine-tuning parameters to function effectively.

\subsection{Future Work}
Future research will focus on extending the VA-IB framework to 3D geometric reasoning by leveraging structured reasoning modules for point clouds~\cite{zhang2026pointcot}. We plan to integrate geometry-aware skip-attention mechanisms~\cite{zhang2026igasa} and investigate coarse-to-fine compression strategies~\cite{zhang2026notallqueries} to further optimize the robustness and efficiency of multimodal reasoning across diverse spatial domains.

\end{document}